\documentclass{elsart}
\usepackage{amssymb}

\renewcommand{\bar}[1]{\overline{#1}}

\usepackage{graphicx}

\providecommand{\Journal}[4] {#1 {\bf #2} (#4) #3}
\providecommand{\LNC}{Lett. Nuovo Cimento } %
\providecommand{\PLB}{Phys. Lett. B } %
\providecommand{\PRL}{Phys. Rev. Lett. } %
\providecommand{\PRD}{Phys. Rev. D } %

\journal{Physics Letters B}

\begin{document}

\begin{frontmatter}
\title{Estimate of neutrino masses from Koide's relation}

\author[pku]{Nan Li},
\author[ccast,pku]{Bo-Qiang Ma\corauthref{cor}}
\corauth[cor]{Corresponding author.} \ead{mabq@phy.pku.edu.cn}
\address[pku]{School of Physics, Peking University, Beijing 100871, China}
\address[ccast]{CCAST (World Laboratory), P.O.~Box 8730, Beijing
100080, China}

\begin{abstract}
We apply Koide's mass relation of charged leptons to neutrinos and
quarks, with both the normal and inverted mass schemes of
neutrinos discussed. We introduce the parameters $k_{\nu}$, $k_u$
and $k_d$ to describe the deviations of neutrinos and quarks from
Koide's relation, and suggest a quark-lepton complementarity of
masses such as $ k_{l}+k_{d} \approx k_{\nu}+k_{u} \approx 2$. The
masses of neutrinos are determined from the improved relation, and
they are strongly hierarchical (with the different orders of
magnitude of $10^{-5}~\mbox{eV}$, $10^{-3}~\mbox{eV}$, and
$10^{-2}~\mbox{eV}$).
\end{abstract}

\end{frontmatter}

\par

{\bf 1. Introduction}

The generation of the masses of fermions is one of the most
fundamental and important problem in theoretical physics. These
masses are taken as free parameters in the standard model of
particle physics and can not be determined by the standard model
itself. Before more underlying theories for this problem to be
found, phenomenological analysis are more useful and practical.
Just like Balmer and Rydberg's formulae for Bohr's theory, several
conjectures for this problem (for example, Barut's
formula~\cite{barut}) have been presented, among which Koide's
relation~\cite{koide,koide2} is one of the most accurate, which
links the masses of charged leptons together,
\begin{equation}\nonumber
m_e+m_{\mu}+m_{\tau}=\frac{2}{3}(\sqrt{m_e}+\sqrt{m_{\mu}}+\sqrt{m_{\tau}})^2,\nonumber
\end{equation}\nonumber
where $m_e$, $m_{\mu}$, $m_{\tau}$ are the masses of electron,
muon, and tau, respectively.

This relation was speculated on the basis of a composite
model~\cite{koide} and the extended technicolor-like
model~\cite{koide2}. The fermion mass matrix in these models is
taken as
\begin{eqnarray}\nonumber
M_f=m_0^fGO_f G,
\end{eqnarray}\nonumber
where $G=\mbox{diag}(g_1,g_2,g_3)$. With the assumptions
$g_i=g^{(1)}+g_{i}^{(8)}$, $\sum_{i}g_{i}^{(8)}=0$ and
$\sum_{i}(g_{i}^{(8)})^2=3(g_{i}^{(1)})^2$, and the charged lepton
mass matrix is the $3\times3$ unit matrix, we can obtain Koide's
relation.

Here we introduce a parameter $k_l$,
\begin{equation}
k_l\equiv\frac{m_e+m_{\mu}+m_{\tau}}{\frac{2}{3}(\sqrt{m_e}+\sqrt{m_{\mu}}+\sqrt{m_{\tau}})^2}.
\end{equation}
With the data of PDG~\cite{PDG},
$m_e=0.510998902\pm0.000000021~\mbox{MeV}$,
$m_{\mu}=105.658357\pm0.000005~\mbox{MeV}$ and
$m_{\tau}=1776.99^{+0.29}_{-0.26}~\mbox{MeV}$, we can get the
range of $k_l=1^{+0.00002635}_{-0.00002021}$, which is perfectly
close to 1.

Foot~\cite{foot} gave a geometrical interpretation for Koide's
relation,
\begin{eqnarray}\nonumber
\cos\theta_l=\frac{(\sqrt{m_e} , \sqrt{m_{\mu}} , \sqrt{m_{\tau}})
\cdot (1,1,1)}{|(\sqrt{m_e} , \sqrt{m_{\mu}} ,
\sqrt{m_{\tau}})||(1,1,1)|}=\frac{\sqrt{m_e}+\sqrt{m_{\mu}}+\sqrt{m_{\tau}}}{\sqrt{3}\sqrt{m_e+m_{\mu}+m_{\tau}}},
\end{eqnarray}\nonumber
where $\theta_l$ is the angle between the points $(\sqrt{m_e} ,
\sqrt{m_{\mu}} , \sqrt{m_{\tau}})$ and $(1,1,1)$. And we can see
that $k_l=\frac{1}{2\cos^2\theta_l}$, and
$\theta_l=\frac{\pi}{4}$.

From the analysis above, we can see the miraculous accuracy of
Koide's relation for charged leptons. A natural question emerges
that whether this excellent relation holds also for neutrinos and
quarks. In Section 2, we apply Koide's relation to neutrinos, with
both the normal and inverted mass schemes considered. In Section
3, we apply Koide's formula to quarks. In Section 4, the masses of
neutrinos are determined by some analogy and conjectures between
leptons and quarks. Finally, in Section 5, we give some discussion
to Koide's relation.

{\bf 2. Koide's relation for neutrinos}

In recent years, the oscillations and mixings of neutrinos have
been strongly established by abundant experimental data. The
long-existed solar neutrino deficit is caused by the oscillation
from $\nu_{e}$ to a mixture of $\nu_{\mu}$ and $\nu_{\tau}$ with a
mixing angle approximately of $\theta_{\mathrm{sol}} \approx
34^{\circ}$ in the KamLAND~\cite{Kam} and SNO~\cite{sno}
experiments. Also, the atmospheric neutrino anomaly is due to the
$\nu_{\mu}$ to $\nu_{\tau}$ oscillation with almost the largest
mixing angle of $\theta_{\mathrm{atm}} \approx 45^{\circ}$ in the
K2K~\cite{K2K} and Super-Kamiokande~\cite{SUPER} experiments.
However, the non-observation of the disappearance of
$\bar{\nu}_{e}$ in the CHOOZ~\cite{Chz} experiment showed that the
mixing angle $\theta_{\mathrm{chz}}$ is smaller than $5^{\circ}$
at the best fit point~\cite{garcia,Altarelli}.

These experiments not only confirmed the oscillations of
neutrinos, but also measured the mass-squared differences of the
neutrino mass eigenstates. According to the global analysis of the
experimental results, we have (the allowed ranges at
$3\sigma$)~\cite{Altarelli}
\begin{equation}
1.4\times10^{-3}~\mbox{eV}^{2}<\Delta
m_{\mathrm{atm}}^{2}=|m_{3}^{2}-m_{2}^{2}|<3.7\times10^{-3}~\mbox{eV}^{2},
\end{equation}
and
\begin{equation}
5.4\times10^{-5}~\mbox{eV}^{2}<\Delta
m_{\mathrm{sol}}^{2}=|m_{2}^{2}-m_{1}^{2}|<9.5\times
10^{-5}~\mbox{eV}^{2},
\end{equation}
where $m_1$, $m_2$, $m_3$ are the masses of the three mass
eigenstates of neutrinos, and the best fit points are
$|m_{3}^{2}-m_{2}^{2}|=2.6\times 10^{-3}~\mbox{eV}^{2}$, and
$|m_{2}^{2}-m_{1}^{2}|=6.9\times 10^{-5}~\mbox{eV}^{2}$
\cite{Altarelli}.

Because of Mikheyev-Smirnov-Wolfenstein~\cite{msw} matter effects
on solar neutrinos, we already know that $m_2>m_1$. Hence we have
\begin{equation}
m_{1}^{2}=m_{2}^{2}-\Delta m_{\mathrm{sol}}^{2},
\end{equation}
and
\begin{equation}
m_{3}^{2}=m_{2}^{2}\pm\Delta m_{\mathrm{atm}}^{2}.
\end{equation}
So there are two mass schemes, (1) the normal mass scheme
$m_3>m_2>m_1$, and (2) the inverted mass scheme $m_2>m_1>m_3$.

Now we will apply Koide's relation to neutrinos. Let us take the
normal mass scheme for example. If Koide's relation holds well for
neutrinos, we have
\begin{equation}
m_1+m_2+m_3=\frac{2}{3}\left(\sqrt{m_1}+\sqrt{m_2}+\sqrt{m_3}\right)^2.
\end{equation}
Substituting Eqs.~(5) and~(6) into Eq.~(1), we get,
\begin{equation}
\sqrt{m^2_2-\Delta m_{\mathrm{sol}}^{2}}+m_2+\sqrt{m^2_2+\Delta
m_{\mathrm{atm}}^{2}}=\frac{2}{3}\left(\sqrt[4]{m^2_2-\Delta
m_{\mathrm{sol}}^{2}}+\sqrt{m_2}+\sqrt[4]{m^2_2+\Delta
m_{\mathrm{atm}}^{2}}\right)^2.
\end{equation}

Solving this equation, we find that there is no real root for
$m_2$ with the restrictions in Eqs.~(3) and~(4). This means that
no matter what value $m_2$ is, Koide's relation does not hold for
neutrinos. So is the inverted mass scheme.

Thus we must improve this relation. Here we introduce a parameter
$k_{\nu}$,
\begin{equation}
k_{\nu}\equiv\frac{m_1+m_2+m_3}{\frac{2}{3}(\sqrt{m_1}+\sqrt{m_2}+\sqrt{m_3})^2}.
\end{equation}
 From the analysis above, we know that $k_{\nu}\not=1$ for
neutrinos. Therefore, only when we have determined the range of
$k_{\nu}$, we can fix the masses of neutrinos. We now check the
situations for the two mass schemes, respectively.

1. For the normal mass scheme, $m_3>m_2>m_1$, we have
\begin{equation}
k_{\nu}=\frac{\sqrt{m^2_2-\Delta
m_{\mathrm{sol}}^{2}}+m_2+\sqrt{m^2_2+\Delta
m_{\mathrm{atm}}^{2}}}{\frac{2}{3}\left(\sqrt[4]{m^2_2-\Delta
m_{\mathrm{sol}}^{2}}+\sqrt{m_2}+\sqrt[4]{m^2_2+\Delta
m_{\mathrm{atm}}^{2}}\right)^2}.
\end{equation}
We can see that $k_{\nu}$ is the function of $m_2$ if $\Delta
m_{\mathrm{sol}}^{2}$ and $\Delta m_{\mathrm{atm}}^{2}$ are fixed.
Due to the inaccuracy of the experimental data, we take $\Delta
m_{\mathrm{sol}}^{2}$ and $\Delta m_{\mathrm{atm}}^{2}$ as their
best fit points here. The range of $k_{\nu}$ is shown in Fig.~1.

\begin{figure}
\begin{center}
\includegraphics[width=7cm]{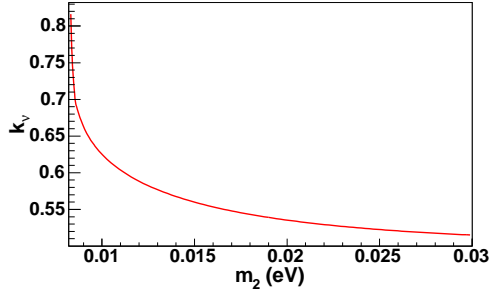}
\caption{The range
of $k_{\nu}$ of the normal mass scheme $m_3>m_2>m_1$.}
\label{Fig1}
\end{center}
\end{figure}

We can see that $0.50<k_{\nu}<0.85$, and $k_{\nu}$ decreases with
the increase of $m_2$. So $k_{\nu}<1$ for neutrinos. This is
different from charged leptons.

2. For the inverted mass scheme, $m_2>m_1>m_3$, we have
\begin{equation}
k_{\nu}=\frac{\sqrt{m^2_2-\Delta
m_{\mathrm{sol}}^{2}}+m_2+\sqrt{m^2_2-\Delta
m_{\mathrm{atm}}^{2}}}{\frac{2}{3}\left(\sqrt[4]{m^2_2-\Delta
m_{\mathrm{sol}}^{2}}+\sqrt{m_2}+\sqrt[4]{m^2_2-\Delta
m_{\mathrm{atm}}^{2}}\right)^2}.
\end{equation}
The range of $k_{\nu}$ is shown in Fig.~2.

\begin{figure}
\begin{center}
\scalebox{0.45}{\includegraphics{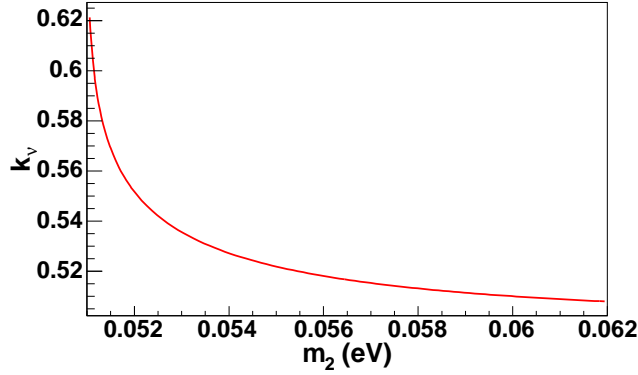}}
\end{center}
\caption{The range of $k_{\nu}$ of the inverted mass scheme
$m_2>m_1>m_3$.}\label{fig2}
\end{figure}

We can see that $0.50<k_{\nu}<0.65$.

Altogether, $0.50<k_{\nu}<0.85$ for both these two mass schemes.
And $k_{\nu}$ of the normal mass scheme is larger than that of the
inverted mass scheme.

{\bf 3. Koide's relation for quarks}

Now we turn to the cases of quarks. Because of the confinement of
quarks, the inaccuracy of the masses of quarks is much bigger than
that of leptons.

Here we take the data of PDG~\cite{PDG}.
\begin{eqnarray}\nonumber
1.5~\mbox{MeV}&<m_u<&4.5~\mbox{MeV},\nonumber\\
1.0~\mbox{GeV}&<m_c<&1.4~\mbox{GeV},\nonumber\\
162.9~\mbox{GeV}&<m_t<&188.5~\mbox{GeV},\\
5~\mbox{MeV}&<m_d<&8.5~\mbox{MeV},\nonumber\\
80~\mbox{MeV}&<m_s<&155~\mbox{MeV},\nonumber\\
4.0~\mbox{GeV}&<m_b<&4.5~\mbox{GeV}.
\end{eqnarray}

1. First, we calculate $k_u$ for $u$, $c$, $t$  quarks, {\it
i.e.}, u-type quarks,
\begin{eqnarray}
k_u&\equiv&\frac{m_u+m_c+m_t}{\frac{2}{3}(\sqrt{m_u}+\sqrt{m_c}+\sqrt{m_t})^2}\nonumber\\
&=&\frac{1+x_u+y_u}{\frac{2}{3}(1+\sqrt{x_u}+\sqrt{y_u})^2},
\end{eqnarray}
where $x_u=m_c/m_u$, $y_u=m_t/m_u$, and we can see that $k_u$ is
the function only of the ratio of the masses of quarks. From
Eq.~(12), we get $2.2\times10^2<x_u<9.3\times10^2$ and
$3.6\times10^4<y_u<1.3\times10^5$. Because Koide's relation is not
energy-scale invariant, the energy scale should be high energy
where the current quark masses rather than the constituent quark
masses should be adopted. The range of $k_u$ is shown in Fig.~3.

\begin{figure}
\begin{center}
\scalebox{0.45}{\includegraphics{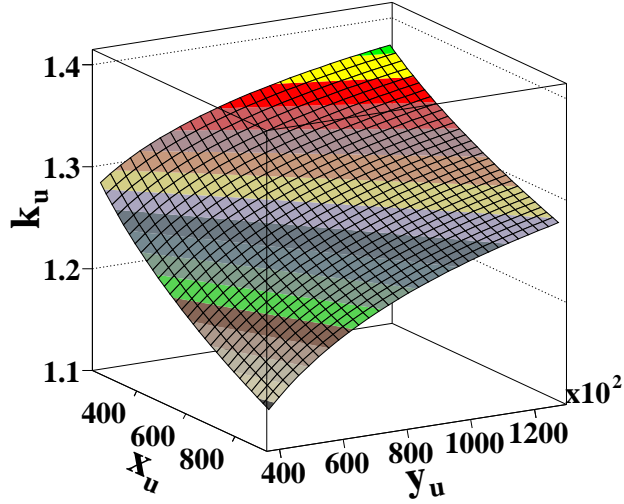}}
\end{center}
\caption{The range of $k_{u}$ for $u$, $c$, $t$
quarks.}\label{Fig3}
\end{figure}

We can see that $1.1<k_u<1.4$. Comparing with the cases of
neutrinos, we find that $k_u>1$ for quarks, and $k_{\nu}<1$ for
neutrinos.

2. Second, we calculate $k_d$ for $d$, $s$, $b$ quarks, {\it
i.e.}, d-type quarks,
\begin{eqnarray}
k_d&\equiv&\frac{m_d+m_s+m_b}{\frac{2}{3}(\sqrt{m_d}+\sqrt{m_s}+\sqrt{m_b})^2}\nonumber\\
&=&\frac{1+x_d+y_d}{\frac{2}{3}(1+\sqrt{x_d}+\sqrt{y_d})^2},
\end{eqnarray}
where $x_d=m_s/m_d$, $y_d=m_b/m_d$. From Eq.~(13), we get
$9.4<x_d<31$ and $4.7\times10^2<y_d<9.0\times10^2$. The range of
$k_d$ is shown in Fig.~4.

\begin{figure}
\begin{center}
\scalebox{0.45}{\includegraphics{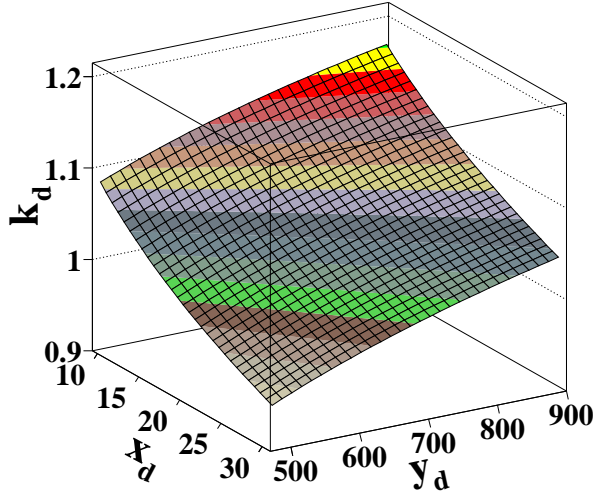}}
\end{center} \caption{The range of $k_{d}$ for $d$,
$s$, $b$ quarks.}\label{Fig4}
\end{figure}

We can see that $0.9<k_d<1.2$. Thus $k_d\approx1$, and this is
similar with the case of charged leptons.

Conclusively, the values of $k_l$, $k_{\nu}$, $k_u$ and $k_d$ can
be summarized as follows
\begin{equation}
 \left(
    \begin {array}{c}
       \nu_{e} \\
       e
\end{array}
\right)
\left(
    \begin {array}{c}
       \nu_{\mu} \\
       \mu
\end{array}
\right)
\left(
    \begin {array}{c}
       \nu_{\tau} \\
       \tau
\end{array}
\right)
\begin {array}{c}
k_{\nu}<1\\
k_{l}=1
\end{array},  \mbox{and}
\left(
    \begin {array}{c}
       u \\
       d
\end{array}
\right)
\left(
    \begin {array}{c}
       c \\
       s
\end{array}
\right)
\left(
    \begin {array}{c}
       t \\
       b
\end{array}
\right)
\begin {array}{c}
k_{u}>1\\
k_{d}\approx1
\end{array}.
\end{equation}

{\bf 4. Estimate of the masses of neutrinos}

We believe that the problem of the generation of the masses of
leptons must be solved together with that of quarks. Since
$k_{l}=1$ and $k_{d}\approx1$, we may conjecture that
$k_{l}+k_{d}\approx 2$. At the same time, since
$0.50<k_{\nu}<0.85$ and $1.1<k_{u}<1.4$, we may analogize the
conjecture of $k_{l}$ and $k_{d}$, and propose the hypothesis that
\begin{equation}
k_{\nu}+k_{u}\approx 2.
\end{equation}
This is from the speculation that there must be some relation
between $k_l$, $k_{\nu}$, $k_u$ and $k_d$. The situation seems to
be similar to the quark-lepton complementarity between mixing
angles of quarks and leptons~\cite{Complementarity}, and we may
call it a quark-lepton complementarity of masses.

Of course, this Ansatz is not the only one of the relations
between $k_l$, $k_{\nu}$, $k_u$ and $k_d$. For example, we may
also assume that $k_l k_d \approx k_{\nu}k_u \approx 1$,
$k_l^2+k_d^2 \approx k_{\nu}^2+k_u^2 \approx 2$, or
$\frac{1}{k_l}+\frac{1}{k_d} \approx
\frac{1}{k_{\nu}}+\frac{1}{k_u} \approx 2$ (this is from the
assumption that $\theta_l+\theta_d \approx \theta_{\nu}+\theta_u
\approx \frac{\pi}{2}$ in Foot's geometrical interpretation).

However, among all of these Ans\"{a}tze, Eq.~(17) is the simplest
one, and it can show the balance between $k_{\nu}$ and $k_u$
(i.e., the quark-lepton complementarity) intuitively and
transparently. Furthermore, the values of $k_{\nu}$ obtained under
other Ans\"{a}tze are close to the value obtained from Eq.~(17),
and the masses of neutrinos are not sensitive to the value of
$k_{\nu}$ (we will show this in the following paragraphs), so we
will use the hypothesis $k_{\nu}+k_{u}\approx 2$ here.

From Fig.~3, we can see that the mean value of $k_{u}$ is 1.25.
Thus from the hypothesis $k_{\nu}+k_{u}\approx2$, we get that
$k_{\nu}\approx 0.75$. This is consistent with the normal mass
scheme and in conflict with the inverted mass scheme. This
indicates that the three masses of neutrinos mass eigenstates are
heavier in order, which is the same as leptons and quarks.

Now we can estimate the absolute masses of neutrinos. Substituting
$k_{\nu}=0.75$, $\Delta m_{\mathrm{atm}}^{2}=2.6\times
10^{-3}~\mbox{eV}^{2}$, and $\Delta m_{\mathrm{sol}}^{2}=6.9\times
10^{-5}~\mbox{eV}^{2}$ into Eq.~(10), we can calculate the value
of $m_2$,
\begin{equation}\nonumber
0.75=\frac{\sqrt{m^2_2-6.9\times
10^{-5}~\mbox{eV}^{2}}+m_2+\sqrt{m^2_2+2.6\times
10^{-3}~\mbox{eV}^{2}}}{\frac{2}{3}\left(\sqrt[4]{m^2_2-6.9\times
10^{-5}~\mbox{eV}^{2}}+\sqrt{m_2}+\sqrt[4]{m^2_2+2.6\times
10^{-3}~\mbox{eV}^{2}}\right)^2},
\end{equation}
and we get $m_2=8.4\times10^{-3}~\mbox{eV}$.

Straightforwardly, we can get
\begin{equation}
m_1=\sqrt{m^2_2-\Delta m_{\mathrm{sol}}^{2}}=1.0 \times
10^{-5}~\mbox{eV},
\end{equation}
and
\begin{equation}
m_3=\sqrt{m^2_2+\Delta m_{\mathrm{atm}}^{2}}=0.05~\mbox{eV}.
\end{equation}
From Eqs.~(18)-(20), we can see that the masses of the neutrino
mass eigenstates are of different orders of magnitude
($10^{-5}~\mbox{eV}$, $10^{-3}~\mbox{eV}$, and
$10^{-2}~\mbox{eV}$), so they are hierarchical, and $m_1$ almost
vanish because $m_2^2$ is very near $\Delta m_{\mathrm{sol}}^{2}$.

Now we can discuss the uncertainty of $m_1$, $m_2$ and $m_3$. In
Fig.~(1), we can see the slope of the curve in very large where
$k_{\nu} \sim 0.75$, so the value of $m_2$ is not sensitive to the
error of $k_{\nu}$. $m_2$ will approximately be
$8.4\times10^{-3}~\mbox{eV}$ even if the mean value of $k_{\nu}$
charges from 0.7 to 0.85, so the value of $m_2$ is precise to a
good degree of accuracy. Similarly, the value of $m_3$ will be
about $0.05~\mbox{eV}$ to a good degree of accuracy too, because
$m_3=\sqrt{m^2_2+\Delta m_{\mathrm{atm}}^{2}}$, and $\Delta
m_{\mathrm{atm}}^{2} \gg m^2_2$. The only point desired to be
mentioned here is the range of $m_1$. Because $m^2_2$ is rather
close to $\Delta m_{\mathrm{sol}}^{2}$, and due to the big
uncertainty of $\Delta m_{\mathrm{sol}}^{2}$, the value of $m_1$
may change largely with $k_{\nu}$. The value $1.0 \times
10^{-5}~\mbox{eV}$ is the rough estimate of the first step, and
its effective number and order of magnitude may change with the
more precise experimental data in the future.

Koide~\cite{koide3} also gave an interpretation of his relation as
a mixing between octet and singlet components in a nonet scheme of
the flavor $U(3)$. He also got the masses of neutrinos
$m_1=0.0026~\mbox{eV}$, $m_2=0.0075~\mbox{eV}$ and
$m_3=0.050~\mbox{eV}$~\cite{koide4}. We can see that his results
are strongly consistent with ours. Especially the values of $m_2$
and $m_3$ are almost the same (only with the exception of $m_1$,
this is because $m^2_2$ is rather close to $\Delta
m_{\mathrm{sol}}^{2}$, and the errors of $\Delta
m_{\mathrm{sol}}^{2}$ is large in nowadays experimental data).

Now we calculate the effective masses of the three flavor
eigenstares of neutrinos, which can be defined as
\begin{equation}
\langle m\rangle_{\alpha}\equiv \sqrt{
\sum\limits_{i=1}^{3}\left(m_i^2|V_{\alpha i}|^{2}\right)},
\end{equation}
where $\alpha=e, \mu, \tau$, and $V_{\alpha i}$ is the element of
the neutrino mixing (MNS) matrix~\cite{mns}, which links the
neutrino flavor eigenstates to the mass eigenstates. The best fit
points of the modulus of MNS matrix are summarized as follows
~\cite{Altarelli}
\begin{equation}
    |V|=\left(
        \begin{array}{ccc}
             0.84 & 0.54 & 0.08\\
             0.44 & 0.56 & 0.71\\
             0.32 & 0.63 & 0.71\\
        \end{array}
        \right).
\end{equation}
Then we get
\begin{eqnarray}
\langle m\rangle_{e}
&=&\sqrt{m_{1}^{2}|V_{e1}|^{2}+m_{2}^{2}|V_{e2}|^{2}+m_{3}^{2}|V_{e3}|^{2}}\nonumber \\
&=&6.0 \times 10^{-3}~\mbox{eV}.
\end{eqnarray}
Similarly,
\begin{equation}
\langle m\rangle_{\mu}=3.6\times 10^{-2}~\mbox{eV},
\end{equation}
\begin{equation}
\langle m\rangle_{\tau}=3.6\times 10^{-2}~\mbox{eV}.
\end{equation}

The upper bounds of $\langle m\rangle_{e}$, $\langle
m\rangle_{\mu}$ and $\langle m\rangle_{\tau}$ are measured by the
experiments $\mbox{H}^3_1 \rightarrow
\mbox{He}^3_2+e+\bar{\nu}_{e}$, $\pi^+ \rightarrow
\mu^++\nu_{\mu}$, and $\tau \rightarrow 5\pi+\nu_{\tau}$,
respectively~\cite{PDG},
\begin{eqnarray}
\langle m\rangle_{e}&<&2.2~\mbox{eV},\nonumber\\
\langle m\rangle_{\mu}&<&0.19~\mbox{MeV}, \nonumber\\
\langle m\rangle_{\tau}&<&18.2~\mbox{MeV}.
\end{eqnarray}
We can see that they are all consistent with the experimental
data, and the more precise planed experiments (for example, KATRIN
experiment~\cite{katrin}) will help to reach a higher sensitivity
to test these results.

Furthermore, we can get the sum of the masses of the neutrino mass
eigenstates,
\begin{equation}
\sum\limits_{i=1}^{3}m_i=0.058~\mbox{eV}.
\end{equation}
This is also consistent with the data from cosmological
observations (Wilkinson microwave anisotropy probe~\cite{wmap} and
2dF Galaxy Redshift Survey~\cite{survey}),
\begin{equation}
\sum\limits_{i=1}^{3}m_i<0.71~\mbox{eV}.
\end{equation}
All the analysis above shows the rationality of our results.

Also, $\langle m\rangle_{\mu}$ and $\langle m\rangle_{\tau}$ are
almost the same because $m_3>m_2>m_1$, and thus the values of
$\langle m\rangle_{\mu}$ and $\langle m\rangle_{\tau}$ are nearly
only dominated by $m_{3}^{2}|V_{\mu3}|^{2}$ and
$m_{3}^{2}|V_{\tau3}|^{2}$. However, $|V_{\mu3}|^{2}\approx
|V_{\tau3}|^{2} \approx 0.71$, so $\langle m\rangle_{\mu}\approx
\langle m\rangle_{\tau}$.

{\bf 5. Summary}

Finally, we give some discussion on our method in determining the
masses of neutrinos. Although the reason and foundation of Koide's
relation is still unknown, there must be some deeper principle
behind this elegant relation, and we believe that this relation
must be applicable to neutrinos and quarks, at least to some
degree. So we introduce the parameters $k_{\nu}$, $k_u$ and $k_d$
to describe the deviations of neutrinos and quarks from Koide's
relation. With this improved relation and the conjecture of a
quark-lepton complementarity of masses such as $k_{l}+k_{d}\approx
k_{\nu}+k_{u}\approx2$, we can determine the absolute masses of
the neutrino mass eigenstates and the effective masses of the
neutrino flavor eigenstates. Due to the inaccuracy of the
experimental data of neutrinos and quarks nowadays, these results
should be only taken as primary estimates. However, if these
results are tested to be consistent with more precise experiments
in the future, it would be a big success of Koide's relation, and
we can get further understanding of the generation of the masses
of leptons and quarks.

{\bf Acknowledgments}

We are very grateful to Prof.~Xiao-Gang He for his stimulating
suggestions and valuable discussions. We also thank Yoshio Koide,
Xun Chen, Xiaorui Lu, and Daxin Zhang for discussions. This work
is partially supported by National Natural Science Foundation of
China and by the Key Grant Project of Chinese Ministry of
Education (NO.~305001).


\end{document}